\documentclass[sigconf]{acmart}
\acmConference[EASE 2026]{The 30th International Conference on Evaluation and Assessment in Software Engineering}{9–12 June, 2026}{Glasgow, Scotland, United Kingdom}
\usepackage{listings}

\usepackage{xcolor}
\copyrightyear{2026}
\acmYear{2026}
\setcopyright{cc}
\setcctype{by}
\acmConference[EASE '26]{International Conference on Evaluation and Assessment in Software Engineering}{June 09--12, 2026}{Glasgow, United Kingdom}
\acmBooktitle{International Conference on Evaluation and Assessment in Software Engineering (EASE '26), June 09--12, 2026, Glasgow, United Kingdom}
\acmDOI{10.1145/3816483.3818228}
\acmISBN{979-8-4007-2348-3/2026/06}

\begin{document}

\title{Evaluating LLM-Based Goal Extraction in Requirements Engineering: Prompting Strategies and Their Limitations}

\author{Anna Arnaudo}
\email{anna.arnaudo@polito.it}
\email{0009-0007-3826-1563}
\affiliation{%
  \institution{Department of Control and Computer Engineering, Politecnico di Torino}
  \city{Torino}
  \country{IT}
}

\author{Riccardo Coppola}
\email{0000-0003-4601-7425}
\affiliation{%
  \institution{Department of Control and Computer Engineering, Politecnico di Torino}
  \city{Torino}
  \country{IT}
}

\author{Maurizio Morisio}
\email{0000-0001-7362-906X}
\affiliation{%
  \institution{Department of Control and Computer Engineering, Politecnico di Torino}
  \city{Torino}
  \country{IT}
}

\author{Flavio Giobergia}
\email{0000-0001-8806-7979}
\affiliation{%
  \institution{Department of Control and Computer Engineering, Politecnico di Torino}
  \city{Torino}
  \country{IT}
}

\author{Andrea Bioddo}
\email{0009-0009-6182-4250}
\affiliation{%
  \institution{Politecnico di Torino}
  \city{Torino}
  \country{IT}
}

\author{Luca Dadone}
\email{0009-0001-2614-2694}
\affiliation{%
  \institution{Politecnico di Torino}
  \city{Torino}
  \country{IT}
}

\author{Angelo Bongiorno}
\email{0009-0007-9639-4583}
\affiliation{%
  \institution{Politecnico di Torino}
  \city{Torino}
  \country{IT}
}

\renewcommand{\shortauthors}{Arnaudo et al.}

\begin{abstract}
Due to the textual and repetitive nature of many Requirements Engineering (RE) artefacts, Large Language Models (LLMs) have proven useful to automate their generation and processing. 
In this paper, we discuss a possible approach for automating the Goal-Oriented Requirements Engineering (GORE) process by extracting functional goals from software documentation through three phases: actor identification, high and low-level goal extraction.
To implement these functionalities, we propose a chain of LLMs fed with engineered prompts. We experimented with different variants of in-context learning and measured the similarities between input data and in-context examples to better investigate their impact. Another key element is the generation-critic mechanism, implemented as a feedback loop involving two LLMs. 
Although the pipeline achieved 61\% accuracy in low-level goal identification — the final stage — these results indicate the approach is best suited as a tool to accelerate manual extraction rather than as a full replacement.
The feedback-loop mechanism with Zero-shot outperformed stand-alone Few-shot, with an ablation study suggesting that performance slightly degrades without the feedback cycle. However, we reported that the combination of the feedback mechanism with Few-shot does not deliver any advantage, possibly suggesting that the primary performance ceiling is the prompting strategy applied to the 'critic' LLM.
Together with the refinement of both the quantity and quality of the Shot examples, future research will integrate Retrieval-Augmented Generation (RAG) and Chain-of-Thought (CoT) prompting to improve accuracy.
\end{abstract}





%\received{}
%\received[revised]{}
%\received[accepted]{}

\maketitle

\section{Introduction}

Rising interest in LLMs for software engineering~\cite{wei2022emergent} highlights their effectiveness in analysing and generating structured artifacts~\cite{akbar2023ethical,kang2023large}. In the field of Requirements Engineering (RE) some studies have explored the extraction of goal models from natural language requirement specifications \cite{das2024extracting}, the extraction of domain models from textual requirements \cite{arulmohan2023extracting}, the enhancement of Use Case definition with LLM-based agents \cite{de2023echo}, goal-model generation from user stories \cite{siddeshwar2024comparative}, class/behavioral model synthesis and benchmarking \cite{bozyigit2024generating}. 

While most current applications of LLMs in software engineering rely on limited interactions with a single model instance, emerging research highlights a shift toward more collaborative paradigms. Integrating Large Language Models into multi-agent systems (MAS) represents a significant advancement, enabling agents to assume specialised roles, coordinate their actions, and collectively address complex software engineering challenges~\cite{feldt2023autonomous,yoon2023autonomous}. Drawing inspiration from the literature, we envision a system architecture organised around multiple LLMs emulating the different steps typically involved in complex processes. 

Moreover, increasing attention is directed towards prompting strategies and their effectiveness in supporting software engineering\cite{khojah_impact_2025}. For this reason, we experimented with different variants of in-context learning to assess how they affect the proposed architecture. 

The ultimate objective of our architecture is the refinement of high-level goals into increasingly concrete, operationalised forms, thus obtaining the list of the system's functional requirements.
Although mapping from low-level goals to API endpoints is not the main focus of this study, we include it in our conceptual architecture because it may provide useful insights about future extensions of this work.

The contributions of this work can be listed as follows:

\begin{itemize}
    \item We proposed a novel architecture that decomposes the GORE process into multiple steps involving a feedback-loop mechanism;
    \item We evaluated the proposed architecture by varying the prompting strategy of the generator model;
    \item We demonstrated that the combination of Zero-shot prompting with a feedback-loop outperforms Few-shot prompting applied to GPT alone in the target tasks;
    \item We measured the similarity between the Few-shot and the ground truth examples, analysing the possible implications on the performances of our system.
\end{itemize}

The replication package of the study is available on Zenodo\footnote{\url{https://zenodo.org/records/18919525}}. 

\subsection{Research Questions (RQs)}

We formalised the evaluation of the proposed architecture through the following RQs:

\begin{itemize}
    \item \textbf{RQ1:} What is the effectiveness of the multi-agent architecture in extracting the list of actors, and how it is influenced by Shot-prompting?
    \item \textbf{RQ2:} What is the effectiveness of the multi-agent architecture in modelling high-level goals, and how it is influenced by Shot-prompting?
    \item \textbf{RQ3:} What is the effectiveness of the multi-agent architecture in decomposing high-level goals into low-level goals, and how it is influenced by Shot-prompting?
    \item \textbf{RQ4:} Does the critic mechanism actually improve the extraction of actors, high and low-level goals?
\end{itemize}

\section{Background}

\subsection{The GORE Framework}

Introduced by Van Lamsweerde in 2001 \cite{van2001goal}, Goal-Oriented Requirements Engineering (GORE) is a technique with the primary objective of identifying all \emph{goals} of a system, defined as \emph{Objectives that the system under consideration should achieve}, which can be then mapped to functional or non-functional requirements. Goals can be formulated at different levels of abstraction, ranging from high-level strategic concerns to low-level technical ones. The technique also includes the identification of the actors --- i.e., stakeholders of the system, including its final users --- goal prioritisation, and conflict detection. 
While low-level goals are typically derived from parent goals by asking ‘how’, the reverse path — goal abstraction — is achieved by answering ‘why’ \cite{van2001goal}.
For sake of simplicity, we consider only functional goals in our work, as the total number of non-functional goals may raise to an unmanageable amount, especially in case of manual datasets curation.

\subsection{Large Language Models in Requirements Engineering}

Zadenoori et al. \cite{zadenoori_large_2025} provide a comprehensive survey of LLM integration into RE, noting an exponential trajectory in publications beginning in 2023. Their analysis reveals that the majority of current studies utilise GPT-family models, with a heavy emphasis on Zero-shot (used in 44\% of the surveyed studies) and Few-shot prompting (29\%). A key finding of the survey is the current inclination towards using models without further task-specific optimisation; the authors argue that this trend highlights a requirement for more rigorous investigation into advanced model architectures and orchestration --- as performed by this work. Moreover, significant hurdles remain concerning the deterministic reliability of these systems and the integration of human supervisory oversight.

Sami et al. \cite{sami_ai_2024} developed a multi-agent system where a \textit{Product Owner} agent generates user stories that are subsequently validated by a \textit{Quality Assurance} agent against the INVEST \cite{alliance_what_2015} and ISO/IEC/IEEE 29148:2018 \footnote{https://www.iso.org/standard/72089.html} standards. Their findings, which address a challenge similar to that addressed herein, emphasise the advantages of inter-agent communication and orchestration over a single, monolithic LLM instance. However, that study only covers the 'user story' format, while in the present work we target system goals.

Das et al. \cite{das2024extracting} propose NLP-driven techniques for extracting structured goal models from unstructured textual input, reducing manual effort and enhancing requirement elicitation accuracy.
Similarly, recent work on API Alignment \cite{feldt2025semantic} integrates GORE with LLM-based techniques by leveraging multiple iterative prompts and the GPT model to extract goals and map them to existing APIs. The paper demonstrates the potential of LLMs for automating goal extraction and API selection, but it also suffers from limitations in validation, including inconsistent goal decomposition, unexplained omissions, and a lack of structured quality control.  

Our approach, building on existing work, aims to expand the investigation of a multi-LLM pipeline for goal extraction and to offer a further exploratory evaluation of its feasibility and limitations.

\section{Approach}
\label{sec:approach}

\subsection{Architecture}
\label{sec:architecture}
We based our architecture on a fixed-structure LLM chain, represented in Figure \ref{fig:architecture}. Since GORE consists of clearly defined and sequential steps, agents' inherent autonomy and flexibility were not well-suited. On the other hand, a structured chain ensures greater control over the execution flow.

At the core of our architecture, two LLMs --- GPT-4 and Llama 3.3 70B --- collaborate through an iterative feedback loop.
GPT-4 functions as the 'generator', while Llama 3.3 70B operates as the 'evaluator'. Although the research prioritises the optimisation of prompting strategies and interaction patterns, this configuration has been chosen since it leverages GPT-4 for high-quality text generation and Llama for a lower-cost response evaluation.
If Llama’s evaluation score is below 8.5/10 --- that will be referred as \emph{Quality Threshold}\footnote{This threshold has been empirically derived, by qualitatively analysing the generated goals.} --- the critique is inserted into the prompt fed to GPT-4 in the next iteration. Until the Quality Threshold is not met, the system can perform a maximum of 3 iterations before continuing to the next phase, as we empirically found that this number is a good trade-off between accuracy and computational overhead. Moreover, we experimentally observed that if the 'generator' and the 'critic' agents fail to reach an agreement within this number of iterations, then it is unlikely to reach convergence. Future works may investigate in more depth the effect on performance when these parameters are varied.

\subsection{Multi-step Pipeline}
\label{sec:pipeline}

\begin{figure*}
    \centering
  \includegraphics[width=0.7\textwidth]{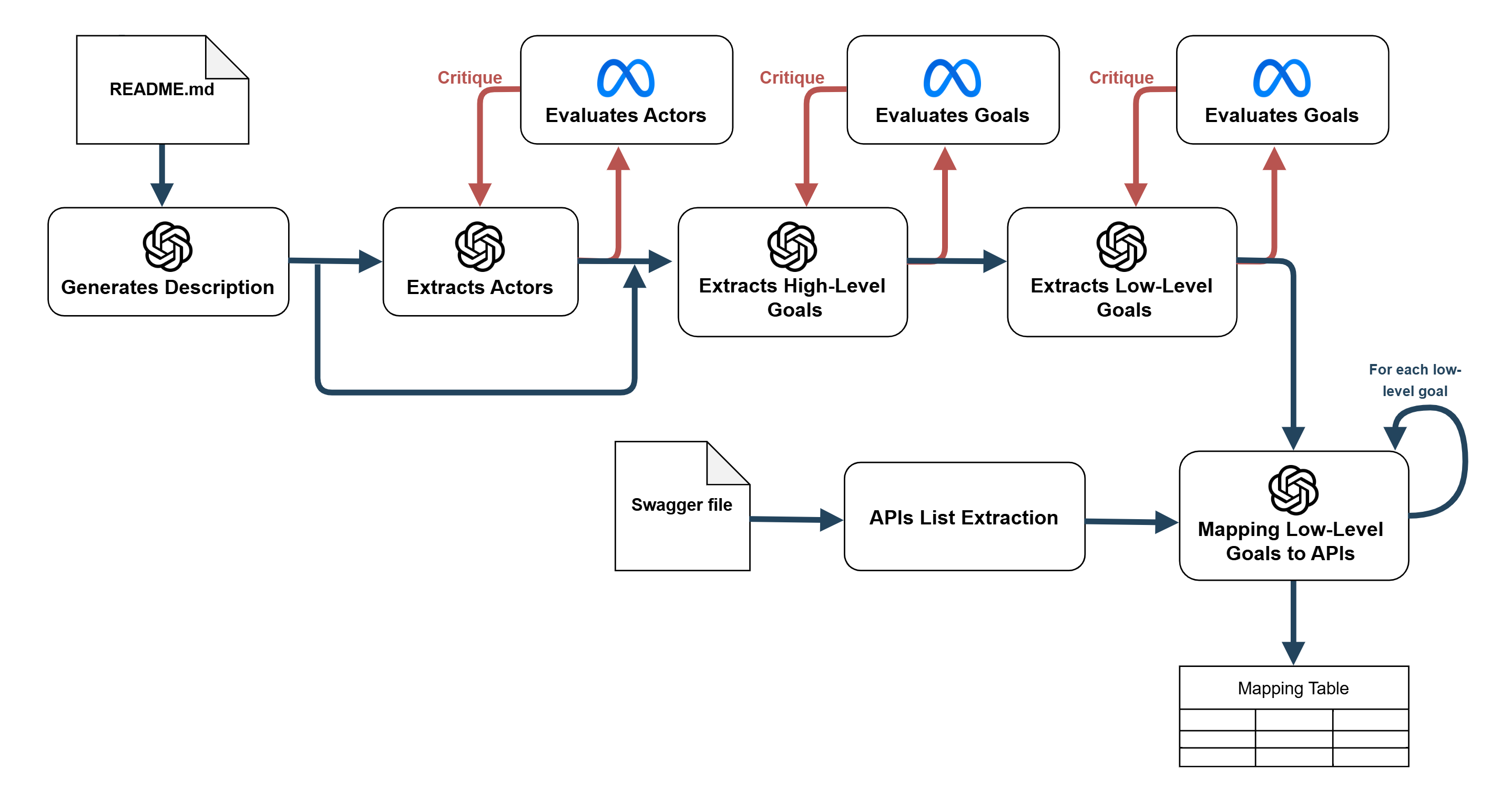}
  \caption{Schema of the proposed architecture, consisting of a LLM chain}
  \Description{}
  \label{fig:architecture}
\end{figure*}

As represented in Figure \ref{fig:architecture}, the pipeline we propose is composed of multiple phases:

\begin{enumerate}
    \item \textbf{Documentation Preprocessing (optional)}: It transforms a raw README file into a natural language project description, serving as a preprocessing step to improve the accuracy of downstream tasks. This is automated through the use of GPT-4, prompted as reported in Table \ref{tab:prompts};
    \item \textbf{Actor Identification}: Actors are active entities that carry actions to achieve one or multiple goals \cite{Kavakli2002GoalOrientedRE};
    \item \textbf{High-level Goals Extraction}: High-level goals are the broad project objectives. For each identified actor, this step aims to extract the main goals that he may want to achieve when interacting with the software;
    \item \textbf{Low-level Goals Extraction}: Low-level goals are specific, actionable objectives. Complex high-level goals are decomposed into a hierarchy of low-level goals;
    \item \textbf{API Mapping}: Goals can be finally mapped to the API endpoints that implement the related functionalities. This step needs as input the set of low-level goals and the API documentation of the software under analysis.
\end{enumerate}

The first step is optional because natural language project descriptions were already available for the \emph{London Ambulance Service} and \emph{Urbain Maintenance} case studies (described in Section \ref{sec:repository}).

Actors, high-level and low-level goals extraction is implemented through the iterative feedback loop described in Section \ref{sec:architecture}, employing the prompts reported in Table \ref{tab:prompts}.

\begin{comment}
\begin{figure}
    \centering
    \includegraphics[width=0.9\columnwidth]{alignment_goals_endpoint_calls.png}
    \caption{Schema representing the hierarchical decomposition of high-level goals.}
    \label{fig:align}
\end{figure}
\end{comment}

\subsection{Prompting Strategy}
\label{sec:prompts}

The chain integrates diverse in-context learning strategies, including Zero-shot, One-shot, and Few-shot. Three Shot examples have been manually curated for each task, by adapting documentation and requirements of existing open-source projects. Specifically, they are related to (i) an application that extracts statistics associated to GitHub accounts; (ii) a food delivery application; and (iii) a home maintenance service locator. Some examples of prompts are reported in Table \ref{tab:prompts}, while Table \ref{tab:shot_examples} contains some instances of Few-shot examples integrated into prompts. The cosine similarities between the ground truth examples --- described in Section \ref{sec:repository} --- and the Shot examples are reported in Table \ref{tab:gpt_shot_similarities}.

\begin{table}
\small
\centering
\caption{Average cosine similarities between the Shot examples used in the prompts for the generator agent --- i.e., GPT --- and the different software projects constituting our ground truth dataset.}
\label{tab:gpt_shot_similarities}
\begin{tabular}{p{0.3\linewidth}p{0.3\linewidth}p{0.2\linewidth}}
\hline
\textbf{Dataset} & \textbf{Task} & \textbf{Average Similarity} \\ \hline
Genome Nexus & Actors & 0.5137 \\
 & high-level goals& 0.5151 \\
 & low-level goals& 0.5323 \\ \hline
Gestao Hospital & Actors & 0.5219 \\
 & high-level goals& 0.5113 \\
 & low-level goals& 0.5082 \\ \hline
London Ambulance & Actors & 0.5319 \\
 & high-level goals& 0.5327 \\
 & low-level goals& 0.5327 \\ \hline
Urban Maintenance & Actors & 0.5007 \\
 & high-level goals& 0.5067 \\
 & low-level goals& 0.5144 \\ \hline
\textbf{Average per Task} & \textbf{Actors} & \textbf{0.5171} \\
 & \textbf{high-level goals} & \textbf{0.5164} \\
 & \textbf{low-level goals} & \textbf{0.5219} \\ \hline
\end{tabular}
\end{table}

While our study evaluates diverse prompting strategies for the 'generator' agent, the 'evaluator' was restricted to a Few-shot configuration to mitigate scoring insensitivity. Empirical observations revealed that --- absent explicit benchmarks for 'major' versus 'minor' errors --- the Llama model lacked the necessary evaluative anchors to assign nuanced ratings. Without these calibrated exemplars, the model exhibited a systemic bias toward invariant scoring across disparate iterations. The cosine similarities between the ground truth examples and the Shot examples feed to the 'evaluator' are reported in Table \ref{tab:llama_shot_similarities}.

\begin{table}
\small
\centering
\caption{Average cosine similarities between the Shot examples fed to the critic agent --- i.e., Llama --- in each task and the different software projects forming the ground truth.}
\label{tab:llama_shot_similarities}
\begin{tabular}{p{0.3\linewidth}p{0.3\linewidth}p{0.2\linewidth}}
\hline
\textbf{Dataset} & \textbf{Task} & \textbf{Average Similarity} \\ \hline
Genome Nexus & Actors & 0.5514 \\
 & high-level goals & 0.5407 \\
 & low-level goals & 0.5257 \\ \hline
Gestao Hospital & Actors & 0.5289 \\
 & high-level goals & 0.4991 \\
 & low-level goals & 0.5066 \\ \hline
London Ambulance & Actors & 0.5031 \\
 & high-level goals & 0.4687 \\
 & low-level goals & 0.5037 \\ \hline
Urban Maintenance & Actors & 0.5382 \\
 & high-level goals & 0.5236 \\
 & low-level goals & 0.5631 \\ \hline
\textbf{Average per Task} & \textbf{Actors} & \textbf{0.5304} \\
 & \textbf{high-level goals} & \textbf{0.5080} \\
 & \textbf{low-level goals} & \textbf{0.5248} \\ \hline
\end{tabular}
\end{table}

At the present stage of development, the framework is restricted to functional requirements by embedding explicit constraints within the model prompts. However, the architecture maintains the flexibility to encompass also non-functional goals. This extension would necessitate removing the existing prompt instructions and updating both the ground-truth data and the Few-shot exemplars to reflect these broader goals.

To enable the reproducibility of the results, the temperature has been set to zero to improve determinism in both GPT and Llama's generations.

Moreover, prompts are enhanced by a priming technique\footnote{In the context of prompt engineering, priming is the practice of strategically providing contextual input within a prompt to shape the model’s responses toward a desired style, reasoning process, or domain of information.}, as can be seen from the prompt samples reported in Table \ref{tab:prompts}.

Finally, it is fundamental to ensure that model outputs are structured and consistent. For this purpose, we use OpenAI’s Pydantic library\footnote{\url{https://ai.pydantic.dev/models/openai/}}, which allows us to enforce a structured schema for GPT-generated responses, while Llama outputs are formatted using standard JSON serialisation.

\begin{table*}[]
    \scriptsize
    \centering
    \caption{Prompts provided to the LLM model to accomplish the different tasks of the proposed pipeline.}
    \label{tab:prompts}
    \begin{tabular}{p{0.07\linewidth}|p{0.83\linewidth}}
\hline
\multicolumn{2}{c}{Preprocess the README file} \\
\hline

\textbf{System prompt:}
&
\begin{lstlisting}
You are a technical writing assistant specialized in summarizing software documentation. Your goal is to extract a 
clear, well-written, and accurate description of a project from its README file. The description should be natural 
and informative, without unnecessary details or implementation specifics. Avoid marketing language, vague claims, 
or filler content. 
Talk as you were a stakeholder describing the system he wants to be implemented (e.g., during requirements elicitation).
\end{lstlisting}
\\
\vspace{0.2cm}

\textbf{Prompt:}
&
\begin{lstlisting}
Here is the README file of a software project: [...]

Explain its purpose, the problem it addresses (if mentioned), and its main functionalities.
\end{lstlisting} \\

    \hline
         \multicolumn{2}{c}{Generate a new list of \textbf{actors}}  \\
         \hline 
         \vspace{0.2cm}
         \textbf{System prompt:} 
         & 
         \scriptsize
\begin{lstlisting}
You are a helpful assistant expert in software engineering tasks, specialised in extracting end-users roles from 
a high level description of a software project. Your task is to extract the actors (roles of end users of the system)
from the given description. If actors are not explicitly mentioned, infer them based on typical users of similar 
software systems. Each extracted actor name should be accompanied by a very short description. 
\end{lstlisting} \\
        \vspace{0.2cm}
        \textbf{Prompt:} 
        & 
        \scriptsize 
\begin{lstlisting} 
Now extract the actors (roles of end users) from the following software description.

**Description:** {description}
\end{lstlisting} \\

        \\

        \hline
        \multicolumn{2}{c}{Generate a new list of \textbf{high-level goals}} \\
        \hline
        \vspace{0.2cm}
        \textbf{System prompt:}
        &
        \scriptsize
        \begin{lstlisting}
You are a helpful assistant expert in software engineering tasks. You're tasked to extract high-level goals 
from a software description for each provided actor that is expected to interact with the software.
Following the Goal-Oriented Requirements Engineering (GORE) frameworks, high-level goals are strategic objectives 
that define the 'why' behind a system. They are usually abstract, business-oriented, and independent of technical 
implementation. They represent the needs of stakeholders or the organization. Focus: Vision and justification.
Generate ONLY the functional goals.
        \end{lstlisting}
        \\
        \vspace{0.2cm}
        \textbf{Prompt:}
        &
        \begin{lstlisting}
Based on your understanding of the typical needs and interests of the following actors in the following 
software project, help generate a list of higl level goals.

**Description:** {project_description}
**Actors:** {actors}
        \end{lstlisting} \\

        \hline
        \multicolumn{2}{c}{
        Generate a new list of \textbf{low-level goals}}
        \\
        \hline
        \vspace{0.2cm}
        \textbf{System prompt:}
        &
        \scriptsize
        \begin{lstlisting}
You are a helpful assistant expert in software engineering tasks. Elicit low-level goals for a 
specific stakeholder in a software project. Avoid generic goals. Instead, break them down into atomic 
actions linked to system capabilities.
Don't be too generic, for example, avoid goals like 'make the software fast', 'develop a web interface' etc.
Following the Goal-Oriented Requirements Engineering (GORE) framework, low-level goals are technical 
objectives that describe 'how' the high-level goals will be achieved. They are more concrete and are eventually 
refined into specific requirements or software specifications. Focus: Implementation and constraints.
Generate ONLY the functional goals.
        \end{lstlisting}
        \\
        \vspace{0.2cm}
        \textbf{Prompt:}
        &
        \scriptsize
        \begin{lstlisting}
Based on your understanding of the typical tasks that compose the following sequence of high-level goals, 
provide if possible a decomposition of goals into sub-goals. Each low-level goal should theoretically correspond 
to a single action of the actor with the software.

**High-level goals:** {highLevelGoals}
        \end{lstlisting} \\

        \hline
         \multicolumn{2}{c}{Critique the response previously generated}  \\
         \hline
         \vspace{0.2cm}
         \textbf{System prompt:} & \scriptsize
         
\begin{lstlisting}
You're an helpful assistant, expert in the field of software engineering.
\end{lstlisting} 
        \\
        \vspace{0.2cm}
         \textbf{Prompt:} 
         &
         \begin{lstlisting}
You're an helpful assistant, expert in the field of software engineering and specialised in the Goal-Oriented 
Requirements Engineering (GORE) framework.
Following the Goal-Oriented Requirements Engineering (GORE) framework: 
-  an actor is active entity that has the capability to perform actions to achieve goals. Unlike goals, which are 
'what' or 'why,' actors are the 'who.'
-  high-level goals are strategic objectives that define the 'why' behind a system. They are usually abstract, 
business-oriented, and independent of technical implementation. They represent the needs of stakeholders or the 
organization. Focus: Vision and justification.
-  low-level goals are technical objectives that describe 'how' the high-level goals will be achieved. They are 
more concrete and are eventually refined into specific requirements or software specifications. Focus: 
Implementation and constraints. 

You can propose new goals taking into account the already present ones. Consider that high-level goals 
often answer the WHY question, while low-level goals often address the HOW.
You must ensure that ONLY functional goals are present.

**Description:** {description}
**Actors:** {actors}
        \end{lstlisting} \\

    \end{tabular}
\end{table*}

\begin{table}[]
    \centering
    \scriptsize
    \caption{Exemplars of Few-shot examples used either for prompting the 'generator' or the 'critic' LLM. Other instances can be found in the online appendix.}
    \label{tab:shot_examples}
    \begin{tabular}{c}
    \toprule
    \textbf{Generation of high-level goals} \\
    \toprule
\begin{lstlisting}
- Description: CatWatch is a web application that tracks and stores
GitHub statistics for accounts. It provides project popularity and 
contributor data through a REST API, offering aggregated stats.
- Actors: [{name: "GitHub account", descr: "Individuals who own 
GitHub accounts"}]

***Output:*** [ "The stakeholder aims to effortlessly monitor the 
popularity metrics of their open source projects across various 
GitHub accounts using CatWatch.", "The stakeholder seeks a feature 
that highlights the most active contributors and collaborators in 
their GitHub repositories through CatWatch.", "The stakeholder 
desires a notification system within CatWatch that alerts them 
promptly about significant activities, such as new contributions 
or rising project trends", "The stakeholder insists on CatWatch 
implementing robust data security measures and compliance with 
privacy standards to safeguard their GitHub account information.",
"The stakeholder requires CatWatch to seamlessly integrate with
their existing workflow tools and development environments,
enhancing productivity and user experience.", "The stakeholder aims
to access detailed analytics and reports generated by CatWatch, 
offering insights into project performance, community engagement,
and other relevant metrics." ]
\end{lstlisting} \\
    
         \toprule
         \textbf{Critic of high-level goals} \\ 
         \toprule
\begin{lstlisting}
- Description: CatWatch is a web application that tracks and stores
GitHub statistics for accounts. It provides project popularity and 
contributor data through a REST API, offering aggregated stats.
- Actors: [{name: "GitHub account", descr: "Individuals who own 
GitHub accounts"}]
- High-level Goals: [ "The stakeholder aims to effortlessly monitor 
the popularity metrics of their open source projects across various
GitHub accounts using CatWatch.", "The stakeholder seeks a feature 
that highlights the most active contributors and collaborators in 
their GitHub repositories through CatWatch.", "The stakeholder wants 
to use the system to track the health, feeding schedules, and GPS 
locations of actual cats in a rescue shelter.", "The stakeholder 
insists on CatWatch implementing robust data security measures and 
compliance with privacy standards to safeguard their GitHub account
information.", "The stakeholder requires CatWatch to seamlessly 
integrate with their existing workflow tools and development 
environments, enhancing productivity and user experience.", "The 
stakeholder aims to access detailed analytics and reports generated 
by CatWatch, offering insights into project performance, community 
engagement, and other relevant metrics." ]

***Score:*** 3/10
***Comment:*** Out of context. Despite the name "CatWatch," the 
goal regarding tracking physical cats is completely unrelated.
\end{lstlisting} \\
\bottomrule
    \end{tabular}
    
\end{table}

\section{Evaluation Method}

\subsection{Evaluation Pipeline}
\label{sec:evaluation_pipeline}
The evaluation procedure consists of the following steps:

\begin{enumerate}
    \item \textbf{Preprocessing}: Stopwords remotion, stemming and lemmatisation are applied to reduce variation in text representation. This can enhance the accuracy of the subsequent similarity computation. Due to their simpler formulation, preprocessing is not applied to actor names, but only on goals descriptions;
    \item \textbf{Encoding}: Both generated and reference data are transformed into vector embeddings using a BERT-based model;
    \item \textbf{Similarity Computation}: Cosine similarity is calculated between generated and reference goal embeddings, forming a similarity matrix;
    \item {\textbf{Maximum Weight Bipartite Matching}}: For each actor or goal, it is necessary to find the element of the ground truth that it aims to resemble. To do so, we employ the algorithm proposed by Munkres et al. \cite{munkres_algorithms_1957} to solve the Maximum Weight Bipartite Matching problem, which prescribes to model the generated and the reference items as nodes of a bipartite graph. In this undirected graph, the arcs are weighted by the cosine similarity between the two strings associated to the nodes. By finding the set of arcs generating the maximum sum of weights, each string is linked to its best match.\footnote{In case the lists of the generated and the reference strings have different lengths, some items remain unpaired. If this happens, those items contribute to the counting of false positives or false negatives respectively.}
    \item \textbf{Computation of Precision, Recall and F1-score}: For the computation of these performance metrics, we employed the formulas from Zhang et al. \cite{zhang_bertscore_2020} reported below:\footnote{They applied these formulas on two sets of words, while we applied them to two sets of sentences.}\\

    \begin{quote}
        $Recall = \frac{1}{|X|} \sum_{(i,j) \in J} \mathbf{x}_i^\top \mathbf{y}_j$ \\

        $Precision = \frac{1}{|Y|} \sum_{(i,j) \in J} \mathbf{x}_i^\top \mathbf{y}_j$ \\

        \textit{F1-score} $= \frac{2*Recall*Precision}{Recall+Precision}$ \\
        
    \end{quote} 

    Where \emph{X} is the set of embeddings from the generated strings; \emph{Y} is the set of embeddings from the ground truth; J is the set of arcs identified in the previous step. The scalar product between $\mathbf{x}_i$ and $\mathbf{y}_i$ computes the cosine similarity between the relative strings.

\end{enumerate}

The approach of computing cosine similarity between BERT embeddings effectively distinguishes true semantic equivalence from surface-level resemblance, ensuring robust adherence to the ground truth. At the same time, solving the Maximum Weight Bipartite Matching problem allows us to effectively measure the similarity between the generated and the reference sets of actors or goals.

\subsection{Experiment Setup}
\label{sec:repository}

To evaluate the proposed architecture, we selected four samples of software projects, as reported in Table \ref{tab:dataset_metadata}. Future work may experiment with more datasets.

We selected two enterprise applications from the WFD (formerly EMB) dataset \cite{andrea_arcuri_webfuzzingemb_2025}. This dataset includes software projects with corresponding READMEs and Swagger-formatted API documentation. Our focus is on GestaoHospital\footnote{\url{https://github.com/ValchanOficial/GestaoHospital}}, a public health management system, and GenomeNexus\footnote{\url{https://github.com/WebFuzzing/Dataset/tree/master/jdk_8_maven/cs/rest-gui/genome-nexus}}, which automates the annotation of cancer-related genetic variants. These projects constitute good examples of commissioned applications involving software engineering activities. 

Furthermore, we examined the \textit{London Ambulance Service} case study, a seminal exemplar frequently cited within the GORE literature \cite{finkelstein_comedy_1996, van_lamsweerde_handling_2000}. The associated annotations were synthesised from the extant body of research.

Finally, we included a software project sourced from an university course, describing an urban maintenance ticketing system. The relative annotations were manually curated by subject-matter experts within the teaching faculty. 

The ground truth was manually curated by three of the authors of this paper, producing the annotations described in Table \ref{tab:dataset_metadata} and available in the replication package.

\begin{table}[]
    \centering
    \small
    \caption{Number of annotations present in the ground truth datasets adopted in this study. HL = High-level, LL = Low-level}
    \label{tab:dataset_metadata}
    \begin{tabular}{lccc}
    \toprule
         \textbf{dataset} & \textbf{Actors} & \textbf{HL Goals} & \textbf{LL Goals} \\
         \toprule
         GestaoHospital & 5 & 4 & 20 \\
         GenomeNexus & 5 & 9 & 34 \\
         Urban Maintenance & 6 & 9 & 18 \\
         London Ambulance System & 4 & 2 & 10 \\ 
         \bottomrule
         \textbf{Total} & \textbf{20} & \textbf{24} & \textbf{82} \\
         \bottomrule
    \end{tabular}
\end{table}

\begin{table}[]
    \small
    \caption{Example of output produced by the proposed system when processing the \textit{Urban Maintenance} use case under the Few-shot setting. Red text represents false positives, while blue highlights false negatives. Examples of false positives are the high-level goals related to account registration and management. While low-level goals have not been reported for brevity, we observed that the relative low-level goals have not been consistently generated. }
    \label{tab:example}
    \centering
    \begin{tabular}{p{0.95\linewidth}}
         \toprule 
         \textbf{Actors} \\
         \toprule
         "Citizens", \textcolor{red}{"Municipal Operators"}, "External Maintenance Personnel", "Administrators", "Non-registered Users", \textcolor{blue}{"Technical Office Staff",
        "Organizational Office Staff"} \\
         \toprule 
         \textbf{High-level goals} \\
         \toprule 
         "Citizens should be able to submit reports by selecting locations on a map and providing necessary details, ensuring their voices are heard in urban management.",
"Municipal operators need tools to review, approve, and manage reports effectively, ensuring timely responses to urban issues.",
"External maintenance personnel should be able to receive assignments, update report statuses, and communicate with municipal operators to resolve issues efficiently.",
"Administrators require access to both public and private statistics to monitor system performance and citizen engagement effectively.",
"Non-registered users should be able to access public reports and statistics to foster trust and engagement with the municipality's initiatives.",
    \textcolor{red}{"The system should allow citizens to submit reports anonymously, ensuring their identity is not disclosed in public listings."}, 
    \textcolor{blue}{"Citizens should be able to create a new account through registration.", "Registered users should be able to manage their account information", "Registered users should be able to start following a report to receive timely notifications", "Administrators should be able to configure the auto-assignment for report categories., "External maintenance users should be able to communicate with the municipal staff, and eventually manage a report's lifecycle through the platform."} \\
    \bottomrule
         
    \end{tabular}
    
\end{table}

\section{Results and Discussion}
\label{sec:results}

All the measurements obtained during the evaluation are arranged in Table \ref{tab:results}, while a qualitative evaluation is reported in Table \ref{tab:example}.

\subsection{Extraction of Actors}
\label{sec:actors}

\begin{table}
    \small
    \centering
    \caption{Results for the tasks described in Section \ref{sec:pipeline}, performed in Zero-Shot (ZS), One-Shot (OS), or Few-shot (FS) setting.}
    \begin{tabular}{p{0.08\linewidth}ccccccccc}
        \toprule
        & \multicolumn{3}{c}{\textbf{Actors}} & \multicolumn{3}{c}{\textbf{High-Level Goals}} & \multicolumn{3}{c}{\textbf{Low-Level Goals}} \\
        \midrule
                  & ZS & OS & FS                          & ZS & OS & FS                            & ZS & OS & FS \\
        \midrule
        Prec.   & 0.75 & 0.68 & \textbf{0.78}  \quad         & \textbf{0.63} & 0.57 & 0.63 \quad            & \textbf{0.78} & 0.72 & 0.77 \\
        Recall      & 0.78 & \textbf{0.80} & 0.67 \quad          & \textbf{0.61} & 0.60 & 0.59 \quad            & \textbf{0.51} & 0.49 & 0.45 \\
        F1    & \textbf{0.76} & 0.74 & 0.72  \quad         & \textbf{0.62} & 0.59 & 0.61  \quad           & \textbf{0.61} & 0.59 & 0.57 \\
        \bottomrule
    \end{tabular}
    \label{tab:results}
\end{table}

As summarised in Table \ref{tab:results}, the Zero-shot setting achieved the best F1-score --- of about 0.76 --- in actor extraction. At the same time, One-shot and Few-shot obtained comparable values for this metric. One-shot is associated with the highest recall (0.80), while Few-shot seem to favour precision --- achieving the peak value of 0.78. This scenario highlights that the absence of Shot examples leads to the best trade-off between precision and recall, but the system's behaviour can be shifted by applying different Shot-prompting strategies.

To assess the impact of the refinement loop described in Section \ref{sec:architecture} --- the core innovation of our architecture --- Table \ref{tab:ablation_llama} reports results from a pipeline using only the GPT model without any feedback mechanism. The comparison reveals that Actor Extraction does not benefit from the feedback loop. Although the observed differences are marginal --- never exceeding three percentage points --- the F1-scores achieved during the ablation study consistently outperform those of the full architecture across all prompting configurations, reaching a peak of 0.78 in the Zero-shot setting.

To further evaluate the impact of the Few-shot prompts, Table \ref{tab:actors_Few-shot_perDataset} details the performance metrics across the ground truth case studies (already described in Section \ref{sec:repository}). Notably, in the \textit{Genome Nexus} case the system achieved perfect precision, albeit with a limited recall of approximately $0.40$. This precision peak aligns with Tables \ref{tab:gpt_shot_similarities} and \ref{tab:llama_shot_similarities}, which identify this dataset as having the highest cosine similarity to the Few-shot exemplars utilised in both GPT and Llama prompts.

Regarding the \textit{Urban Maintenance} dataset --- which yielded the second-highest precision and a substantial recall of $0.70$ --- the average cosine similarity does not rank second for GPT prompts; however, it does hold the second-highest position for Llama-based prompts (Table \ref{tab:llama_shot_similarities}). A consistent pattern emerges for the \textit{Gestao Hospital} and \textit{London Ambulance Service} datasets, which rank third and fourth in precision, respectively. Their precision scores correlate proportionally with the average cosine similarities observed in the Llama prompts. This suggests that the order of precision is preserved specifically in relation to the Llama Few-shot similarities, potentially indicating a direct correlation between the diversity of exemplars provided to the critic agent and the system's overall precision.

However, analogous patterns cannot be found when considering neither system's recall nor F1-score. Similarly, the cosine similarities between the case studies forming the ground truth and the Few-Shot examples used in GPT prompting seem not to be involved. 
\\[1ex]

\begin{table}[h]
\small
\centering
\caption{Performance metrics for the Actor identification task using Few-shot prompting, disaggregated by the individual software projects within the ground truth and ordered by precision.}
\label{tab:actors_Few-shot_perDataset}
\begin{tabular}{lccc}
\toprule
\textbf{Dataset} & \textbf{Recall} & \textbf{Precision} & \textbf{F1-score} \\ \hline
Genome Nexus & 0.40 & 1.00 & 0.57 \\
Urban Maintenance & 0.70 & 0.84 & 0.77 \\
London Ambulance Service & 0.76 & 0.51 & 0.61 \\
Gestao Hospital & 0.85 & 0.71 & 0.77 \\
 \bottomrule
\end{tabular}
\end{table}

\begin{table}
    \centering
    \small
    \caption{Results for the Llama's feedback ablation, performed in Zero-Shot (ZS), One-Shot (OS), or Few-shot (FS) setting.}
    \begin{tabular}{p{0.08\linewidth}ccccccccc}
        \toprule
        & \multicolumn{3}{c}{\textbf{Actors}} & \multicolumn{3}{c}{\textbf{High-Level Goals}} & \multicolumn{3}{c}{\textbf{Low-Level Goals}} \\
        \midrule
                  & ZS & OS & FS &                            ZS & OS & FS &                      ZS & OS & FS \\
        \midrule
        Prec.                   & 0.80 & 0.80 & \textbf{0.86}              & 0.65 & 0.66 & \textbf{0.68}                & 0.79 & 0.76 & \textbf{0.79} \\
        Recall                      & \textbf{0.77} & 0.74 & 0.67               & 0.46 & 0.46 & \textbf{0.53}       & 0.38 & 0.36 & \textbf{0.40} \\
        F1                    & \textbf{0.78} & 0.77 & 0.75               & 0.54 & 0.54 & \textbf{0.60}       & 0.51 & 0.50 & \textbf{0.53} \\
        \bottomrule
    \end{tabular}
    \label{tab:ablation_llama}
\end{table}

\noindent\fbox{%
    \parbox{\columnwidth}{%
        \textbf{Answer to RQ1:} Our evaluation reported an F1-Score of 0.76 in the optimal prompting configuration — specifically Zero-shot. This indicates that our architecture takes limited benefit from in-context learning. Furthermore, a distinct trade-off is evident between the maximum precision of 0.78 achieved via Few-shot and the maximum recall of 0.80 associated with One-shot prompting. The Llama ablation study yielded a slightly superior F1-score of 0.78 --- associated with Zero-shot again. This suggests that the generator agent --- i.e., GPT --- is the primary responsible of the suboptimal exploitation of the Shot examples. Finally, we found a possible correlation between the precision in Actor extraction and the cosine similarity between the software project's description and the Few-Shot examples provided to the critic agent --- i.e., the Llama model. This correlation suggests that the Few-shot strategy applied to Llama should be enriched to improve the critic mechanism's effectiveness.
    }
}

\subsection{Extraction of High-Level Goals}
\label{sec:hl}

As reported in Table \ref{tab:results}, high-level goal extraction demonstrates a distinct behavioural pattern across prompting settings compared to actor extraction. The Zero-shot strategy yielded the optimal overall F1-score, precision, and recall — recorded as 0.62, 0.63, and 0.61, respectively. These findings confirm the detrimental impact of Shot-prompting if integrated with the feedback mechanism. 

However, in the results of the Llama ablation study --- reported in Table \ref{tab:ablation_llama} --- the opposite trend can be observed, with the supremacy of Few-shot prompting. This may suggest that --- for what concerns the task of identifying high-level goals --- the GPT model takes advantage from Shot examples, but this effect is neutralised by the critique mechanism. 

In high-level goals extraction, the contribution of Llama proves to be slightly beneficial: during the ablation study, the maximum F1-score reached was 0.60, which is two points below the one relative to the complete architecture.

These results may indicate that, while Few-shot prompting improves the 'generator' model alone, high-level goals extraction takes greater advantage by the introduction of the feedback mechanism. 

Finally, in Table \ref{tab:hl_Few-shot_perDataset} we report the performance achieved by the complete architecture divided by case study, when Few-shot is applied. As was observed for the actors extraction task, a proportional relation --- albeit not linear --- can be found when comparing the precision metric achieved in each case study and the cosine similarities reported in Table \ref{tab:llama_shot_similarities} --- which have been computed between the case studies documentation and the few Shot examples fed to the Llama model. This may suggest the primary bottleneck resided in the quality of the Few-Shot examples provided to the 'critic' agent, and not to the 'generator' model. 

Consistently with the results presented in Section \ref{sec:actors}, we did not observe any correlation between the similarity to the GPT's Few-shot examples of each case study and the respective performance achieved by the system.
\\[1ex] 

\begin{table}[h]
\small
\centering
\caption{Performance metrics for the high-level goals identification task using Few-shot prompting, disaggregated by the individual software projects within the ground truth and ordered by precision.}
\label{tab:hl_Few-shot_perDataset}
\begin{tabular}{lccc}
\toprule
\textbf{Dataset} & \textbf{Recall} & \textbf{Precision} & \textbf{F1-score} \\ \hline
Genome Nexus & 0.51 & 0.76 & 0.61 \\
London Ambulance Service & 0.75 & 0.21 & 0.33 \\
Urban Maintenance & 0.53 & 0.69 & 0.60 \\
Gestao Hospital & 0.81 & 0.46 & 0.59 \\
\bottomrule
\end{tabular}
\end{table}

\noindent\fbox{%
    \parbox{\columnwidth}{%
        \textbf{Answer to RQ2:} The maximum F1-score reached by our architecture for the extraction of high-level goals is 0.62 (Zero-shot). The ablation of the critique mechanism reveals an opposite trend, with Few-shot achieving the highest F1-score --- which was 0.60. This may suggest that the performance improvements achieved by the addition of the feedback loop overcome the ones related to the application of Few-shot to the GPT model alone. The performance with the feedback mechanism during Few-shot learning shows a specific trend: precision correlates proportionally --- though not linearly --- with the cosine similarity between the case studies and the examples provided to the Llama model. This pattern consistently mirrors the results previously observed during the actors extraction task, possibly indicating that the primary bottleneck resided in the quality of the in-context examples provided to the 'critic' model.
    }
}

\subsection{Extraction of Low-Level Goals}

As shown in Table \ref{tab:results}, low-level Goal extraction yields a F1-score slightly lower than high-level goals extraction (0.61 versus 0.62). This can indicate that the errors may have propagated up to this point in the pipeline, effectively establishing a performance ceiling. Indeed, as detailed below, it is possible to confirm some patterns already found when analysing low-level goals extraction in Section \ref{sec:hl}.

In the Zero-Shot setting, the system achieves the best recall (0.78), precision (0.51), and F1-score (0.61). At the same time, Few-shot proves to be the best strategy in the absence of the feedback mechanism: in the results of the Llama's ablation study --- reported in Table \ref{tab:ablation_llama} --- Few-shot is associated with a great precision (0.78), a low recall (0.40), and the highest F1-score (0.53). These values confirm the phenomena already observed in the extraction of high-level goals, suggesting that the introduction of the feedback mechanism yields greater benefits than the application of Few-shot prompting to the GPT model alone. The iterative interaction between GPT and Llama failed to exploit Shot-prompting, which suggests that the primary bottleneck resides within the prompting of the Llama model.

Following the established methodology, we compared the cosine similarities between the case studies and the Llama's Few-shot exemplars — detailed in Table \ref{tab:llama_shot_similarities} — against the performance metrics achieved by the architecture in the Few-shot setting for each corresponding case study (Table \ref{tab:ll_fewshot_metrics}). Differently from the previous two sections, we were unable to find any correlation pattern. This may suggest that the errors made during the extraction of high-level goals may have introduced perturbations in this final step. As can be seen by the prompts in Table \ref{tab:prompts}, the models are tasked to extract low-level goals by starting only from the high-level ones. Future work may investigate this effect more deeply by isolating the generation of low-level goals. \\[1ex]

\begin{table}[h]
\small
\centering
\caption{Performance metrics for the Low-level goals identification task using Few-shot prompting, disaggregated by the individual software projects within the ground truth and ordered by precision.}
\label{tab:ll_fewshot_metrics}
\begin{tabular}{lccc}
\toprule
\textbf{Dataset} & \textbf{Recall} & \textbf{Precision} & \textbf{F1-score} \\ \hline
Genome Nexus & 0.25 & 0.85 & 0.39 \\
Gestao Hospital & 0.56 & 0.76 & 0.64 \\
Urban Maintenance & 0.61 & 0.69 & 0.65 \\
London Ambulance Service & 0.68 & 0.68 & 0.68 \\
\bottomrule
\end{tabular}
\end{table}

\noindent\fbox{%
    \parbox{\columnwidth}{%
        \textbf{Answer to RQ3:}
        The maximum F1-Score obtained for low-Level Goals extraction is 0.61 (Zero-shot). This value is slightly lower than the one achieved for the previous task in the pipeline, suggesting that the propagation of errors may have introduced performance ceiling. By removing the feedback loop from the architecture, the system achieves a maximum F1-score of 0.53 in the Few-shot setting. While this indicates that the introduction of the critic agent delivers a tangible advantage, it confirms that this benefit comes at the cost of neutralising the effects of in-context, as this phenomenon was already observed in the high-level goals extraction.
    }%
}

\subsection{API Mapping}

\noindent
\begin{table*}
\centering
    \footnotesize
        \caption{Examples of generated API mappings}

    \begin{tabular}
    {p{0.2\textwidth}p{0.55\textwidth}p{0.15\textwidth}}
    \toprule
    \textbf{High Level Goal Name} & \textbf{Low Level Goal Name} & \textbf{API Name} \\
    \hline
Manage Healthcare Operations & Register a new hospital with essential details, including name, address, and contact information. & insertUsingPOST \\
Manage Healthcare Operations & Retrieve a list of all registered hospitals with their registration status and details. & findAllUsingGET \\
\bottomrule
    \end{tabular}
    \label{tab:apis}
\end{table*}

Given the limited F1-scores observed in the preceding phases, this component is considered exploratory rather than part of the formal evaluation. In Table~\ref{tab:apis}, we therefore provide only a qualitative illustration of how such a mapping might look when applied to the extracted goals.
These mappings should be interpreted as plausible candidates generated by the model rather than validated correspondences. Nevertheless, following a manual review of the project’s API documentation, the authors regard them as substantively sound.
This provides preliminary insight into how such a component could support analysts during design or requirements traceability tasks.

\section{Limitations}

\iffalse    %troppo
\begin{table}
    \footnotesize
    \centering
    \caption{Quality concepts for assessing a study's validity \cite{kitchenham_guidelines_2007}.}
    \label{tab:quality_concepts}
    \begin{tabular}{|p{0.25\linewidth}|p{0.65\linewidth}|}
        \hline
         \textbf{Name} & \textbf{Definition}\\
         \hline
         Bias & A tendency to produce results that depart systematically from the ‘true’ results. \\
         \hline
         Internal Validity & The extent to which the design and conduct of the study are likely to prevent systematic error. \\
         \hline
         External Validity & The extent to which the effects observed in the study are applicable outside of the study. \\
         \hline
    \end{tabular}

\end{table}
\fi

\subsection{Bias}
Some bias may have been introduced through the examples employed in the in-context learning. Specifically, if the examples used were more closely aligned with the target problem, the results could have been biased towards higher performance without actually being the consequence of better design choices. This could be mitigated by expanding the evaluation to further benchmarks, or by introducing RAG-augmented Few-shot prompting.\footnote{Retrieval Augmented Generation (RAG) with Few-shot prompting refers to the use of a RAG system to retrieve the most pertinent Few-Shot examples at inference time, based on their similarity with the system's current input.}

Moreover, a potential source of bias lies in the input software documents (Section \ref{sec:pipeline}). As highlighted in the prompt engineering literature \cite{lin_how_2023, errica_what_2025}, the clarity and quality of information contained within input prompts are crucial determinants, and the adopted preprocessing strategy may not be enough for bare README files. Future work may assess the impact of feeding the system with more extensive documentation. 

\subsection{Threats to Construct Validity}

In the present study, README files have been used as a proxy for preliminary documents from which the requirements might be elicited. To provide more realistic application scenarios, the approach should be evaluated with natural language requirements (e.g., transcripts of interviews with stakeholders) rather than README files. 

Furthermore, the multi-step pipeline enforces a 'waterfall' methodology that precludes the modification of high-level goals during the elicitation of low-level objectives. This contrasts with the established literature \cite{van2001goal}, which advocates for the late-stage discovery of high-level goals through obstacle analysis and by addressing 'why' queries relative to low-level goals. Future work may enhance architectural flexibility by integrating these additional mechanisms.

\subsection{Threats to Internal Validity}

We acknowledge that our study did not answer the proposed research questions exhaustively. Indeed, more experiments could be performed by varying the values of the \emph{Quality Threshold} and the maximum number of iterations of the generation-critique loop (both defined in Section \ref{sec:architecture}).

Furthermore, our method stipulates that the final response produced by GPT-4 is retained once the maximum number of iterations has been reached. Although retaining the final iteration may be suboptimal compared to selecting the highest-scoring response, Llama’s feedback effectively mitigates quality degradation across successive outputs.

\subsection{Threats to External Validity}

As previously stated, we concentrated solely on functional goals. While this limitation was imposed to keep the number of goals per use case manageable for manual annotation, we acknowledge that it may limit the generalisability of our results to real-world scenarios where non-functional goals are critical. 

Finally, it is worth noting that the reported results are highly dependent on the chosen architecture, the specific models employed, and the datasets used to validate the approach. Although they offer valuable insights into the capabilities of LLMs, these findings may not be generalisable to other combinations of models or alternative configurations (e.g., changing the number of max iterations) within a processing chain.

\section{Conclusion and Future Work}
We presented a semi-structured approach for automating parts of the RE process, assessing the impact of in-context learning, and exploiting models with diverging base knowledge for refining the outputs through an iterative feedback mechanism.
However, the results were not entirely satisfactory, with an F1-score of 61\% in low-level goals extraction --- the last step of the pipeline. We acknowledge that the values observed are insufficient for fully automated use and would still require substantial manual supervision in practice. We therefore view the proposed approach not as a replacement for human annotation, but as a starting point that can assist and accelerate manual extraction. Improving recall is a key direction for future work, and systematic comparison with human recall on the same data would provide a more meaningful upper bound and evaluation target.

Crucially, the ablation study  —  conducted by removing the Llama-based feedback mechanism  —  demonstrated that the proposed architecture yields consistent advantages over a conventional, linear pipeline utilising GPT in isolation. Moreover, we were able to demonstrate that our feedback-loop mechanism with Zero-shot prompting outperforms Few-shot prompting applied to the GPT model alone.

Measurements of the cosine similarity between the case study descriptions and the shot examples provided to the GPT model revealed no correlation with performance per case study. This suggests a balanced variety among the in-context examples.

Conversely, analysis of the Few-shot examples used for Llama prompting indicated a potential correlation between their similarity to the case studies and the precision values achieved. This suggests that the quantity and variety of in-context examples provided to the Llama model should be expanded to improve robustness.

While our approach may serve as an encouraging starting point for developing more accurate systems, we recognise that there remains substantial room for improvement.
Future work may involve conducting repeated runs to assess stability, evaluating additional datasets (Section \ref{sec:repository}), and testing alternative embedding models --- such as Sentence BERT --- in place of BERT (Section \ref{sec:evaluation_pipeline}). 

In its current form, our method - although iterative - did not involve a human in the loop, which may account for the suboptimal outcomes. Indeed, various studies in the RE literature \cite{ferrari_formal_2025, ebrahim_enhancing_2025, vogelsang_prompting_2024} indicate that reliable results are difficult to achieve without involving humans.
Furthermore, Llama’s evaluation process could be enhanced with an increased number of Few-Shot examples, which are fundamental to provide to the model the references to assign the scores. Moreover, the model could be provided with literature-grounded instructions - drawing from RE and GORE studies - to guide its outputs towards a more informed evaluation, rather than depending exclusively on its internal knowledge and probabilistic reasoning. This enhancement could be realised through the integration of a Retrieval-Augmented Generation (RAG) system and the application of Chain of Thought (CoT) prompting.

Additionally, our architecture could be modified to better align to the GORE procedures described in the literature. Specifically, a further loop mechanism should be introduced, encompassing both the high-level and low-level goals extraction phases to enable late discovery of high-level goals.

Finally, our study was conducted without imposing constraints on computational resources or processing time. Although we relied on remote API calls to access both the GPT and Llama models, future research could examine in greater depth the computational and economic costs.

\bibliographystyle{ACM-Reference-Format}
\bibliography{bib}

@article{khojah_impact_2025,
	title = {The Impact of Prompt Programming on Function-Level Code Generation},
	volume = {51},
	rights = {https://creativecommons.org/licenses/by/4.0/legalcode},
	issn = {0098-5589, 1939-3520, 2326-3881},
	url = {https://ieeexplore.ieee.org/document/11077752/},
	doi = {10.1109/TSE.2025.3587794},
	pages = {2381--2395},
	number = {8},
	journal = {{IEEE} Transactions on Software Engineering},
	shortjournal = {{IIEEE} Trans. Software Eng.},
	author = {Khojah, Ranim and De Oliveira Neto, Francisco Gomes and Mohamad, Mazen and Leitner, Philipp},
	urldate = {2026-05-23},
	year = {2025},
}

@article{Kavakli2002GoalOrientedRE,
  title={Goal-Oriented Requirements Engineering: A Unifying Framework},
  author={Evangelia Kavakli},
  journal={Requirements Engineering},
  year={2002},
  volume={6},
  pages={237-251},
  url={https://api.semanticscholar.org/CorpusID:17959073}
}

@misc{finkelstein_comedy_1996,
	location = {Schloss Velen, Germany},
	title = {A comedy of errors: the London Ambulance Service case study},
	isbn = {978-0-8186-7361-0},
	url = {http://ieeexplore.ieee.org/document/501141/},
	doi = {10.1109/IWSSD.1996.501141},
	shorttitle = {A comedy of errors},
	eventtitle = {8th International Workshop on Software Specification and Design},
	pages = {2--4},
	booktitle = {Proceedings of the 8th International Workshop on Software Specification and Design},
	publisher = {{IEEE} Comput. Soc. Press},
	author = {Finkelstein, A. and Dowell, J.},
	urldate = {2026-02-28},
	year = {1996},
}

@article{van_lamsweerde_handling_2000,
	title = {Handling obstacles in goal-oriented requirements engineering},
	volume = {26},
	rights = {https://ieeexplore.ieee.org/Xplorehelp/downloads/license-information/{IEEE}.html},
	issn = {0098-5589, 1939-3520, 2326-3881},
	url = {https://ieeexplore.ieee.org/document/879820/},
	doi = {10.1109/32.879820},
	pages = {978--1005},
	number = {10},
	journal = {{IEEE} Transactions on Software Engineering},
	shortjournal = {{IIEEE} Trans. Software Eng.},
	author = {Van Lamsweerde, A. and Letier, E.},
	urldate = {2026-02-28},
	year = {2000-10},
}

@misc{zhang_bertscore_2020,
	title = {{BERTScore}: Evaluating Text Generation with {BERT}},
	url = {http://arxiv.org/abs/1904.09675},
	doi = {10.48550/arXiv.1904.09675},
	shorttitle = {{BERTScore}},
	number = {{arXiv}:1904.09675},
	publisher = {{arXiv}},
	author = {Zhang, Tianyi and Kishore, Varsha and Wu, Felix and Weinberger, Kilian Q. and Artzi, Yoav},
	urldate = {2026-02-28},
	year = {2020-02-24},
	eprinttype = {arxiv},
	eprint = {1904.09675 [cs]},
}

@article{munkres_algorithms_1957,
	title = {Algorithms for the Assignment and Transportation Problems},
	volume = {5},
	issn = {0368-4245, 2168-3484},
	url = {http://epubs.siam.org/doi/10.1137/0105003},
	doi = {10.1137/0105003},
	pages = {32--38},
	number = {1},
	journal = {Journal of the Society for Industrial and Applied Mathematics},
	shortjournal = {Journal of the Society for Industrial and Applied Mathematics},
	author = {Munkres, James},
	urldate = {2026-02-28},
	year = {1957},
	langid = {english},
}

@misc{vogelsang_prompting_2024,
    year = {2024},
	title = {Prompting the Future: Integrating Generative {LLMs} and Requirements Engineering},
	author = {Vogelsang, Andreas},
	langid = {english},
}

@misc{ebrahim_enhancing_2025,
year = {2025}, 
	location = {Vienna, Austria},
	title = {Enhancing Software Requirements Engineering with Language Models and Prompting Techniques: Insights from the Current Research and Future Directions},
	isbn = {979-8-89176-254-1},
	url = {https://aclanthology.org/2025.acl-srw.31/},
	doi = {10.18653/v1/2025.acl-srw.31},
	shorttitle = {Enhancing Software Requirements Engineering with Language Models and Prompting Techniques},
	pages = {486--496},
	booktitle = {Proceedings of the 63rd Annual Meeting of the Association for Computational Linguistics (Volume 4: Student Research Workshop)},
	publisher = {Association for Computational Linguistics},
	author = {Ebrahim, Moemen and Guirguis, Shawkat and Basta, Christine},
	editor = {Zhao, Jin and Wang, Mingyang and Liu, Zhu},
	urldate = {2025-11-09},
	date = {2025},
}

@misc{ferrari_formal_2025,
year = {2025},
	title = {Formal requirements engineering and large language models: A two-way roadmap},
	volume = {181},
	issn = {0950-5849},
	url = {https://doi.org/10.1016/j.infsof.2025.107697},
	doi = {10.1016/j.infsof.2025.107697},
	shorttitle = {Formal requirements engineering and large language models},
	issue = {C},
	journal = {Inf. Softw. Technol.},
	author = {Ferrari, Alessio and Spoletini, Paola},
	urldate = {2025-11-09},
}

@misc{kitchenham_guidelines_2007,
year = {2007},
	title = {Guidelines for performing systematic literature reviews in software engineering},
	url = {https://docs.edtechhub.org/lib/EDAG684W},
	institution = {Technical report, {EBSE} Technical Report {EBSE}-2007-01},
	author = {Kitchenham, B. and Charters, S.},
	urlyear = {2025-04-12},
	year = {2007},
	langid = {british},
}

@misc{wei2022emergent,
  title={Emergent abilities of large language models},
  author={Wei, Jason and Tay, Yi and Bommasani, Rishi and Raffel, Colin and Zoph, Barret and Borgeaud, Sebastian and Yogatama, Dani and Bosma, Maarten and Zhou, Denny and Metzler, Donald and others},
  journal={arXiv preprint arXiv:2206.07682},
  year={2022}
}

@misc{van2001goal,
  title={Goal-oriented requirements engineering: A guided tour},
  author={Van Lamsweerde, Axel},
  booktitle={Proceedings fifth ieee international symposium on requirements engineering},
  pages={249--262},
  year={2001},
  publisher={IEEE}
}

@misc{kang2023large,
  title={Large language models are few-shot testers: Exploring llm-based general bug reproduction},
  author={Kang, Sungmin and Yoon, Juyeon and Yoo, Shin},
  booktitle={2023 IEEE/ACM 45th International Conference on Software Engineering (ICSE)},
  pages={2312--2323},
  year={2023},
  publisher={IEEE}
}

@misc{feldt2025semantic,
  title={Semantic API Alignment: Linking High-level User Goals to APIs},
  author={Feldt, Robert and Coppola, Riccardo},
  booktitle={2025 IEEE/ACM International Workshop on Natural Language-Based Software Engineering (NLBSE)},
  pages={17--20},
  year={2025},
  publisher={IEEE}
}

@misc{yoon2023autonomous,
      title={Autonomous Large Language Model Agents Enabling Intent-Driven Mobile GUI Testing}, 
      author={Juyeon Yoon and Robert Feldt and Shin Yoo},
      year={2023},
      eprint={2311.08649},
      archivePrefix={arXiv},
      primaryClass={cs.SE}
}

@misc{feldt2023autonomous,
      title={Towards Autonomous Testing Agents via Conversational Large Language Models}, 
      author={Robert Feldt and Sungmin Kang and Juyeon Yoon and Shin Yoo},
      year={2023},
      eprint={2306.05152},
      archivePrefix={arXiv},
      primaryClass={cs.SE}
}

@misc{de2023echo,
  title={Echo: An approach to enhance use case quality exploiting large language models},
  author={De Vito, Gabriele and Palomba, Fabio and Gravino, Carmine and Di Martino, Sergio and Ferrucci, Filomena},
  booktitle={2023 49th Euromicro Conference on Software Engineering and Advanced Applications (SEAA)},
  pages={53--60},
  year={2023},
  publisher={IEEE}
}

@misc{arulmohan2023extracting,
  author={Arulmohan, Sathurshan and Meurs, Marie-Jean and Mosser, Sébastien},
  booktitle={2023 ACM/IEEE International Conference on Model Driven Engineering Languages and Systems Companion (MODELS-C)}, 
  title={Extracting Domain Models from Textual Requirements in the Era of Large Language Models}, 
  year={2023},
  volume={},
  number={},
  pages={580-587},
  doi={10.1109/MODELS-C59198.2023.00096},
  publisher={IEEE}
}

@article{das2024extracting,
    year={2024},
	title = {Extracting goal models from natural language requirement specifications},
	volume = {211},
	issn = {0164-1212},
	url = {https://www.sciencedirect.com/science/article/pii/S0164121224000244},
	doi = {https://doi.org/10.1016/j.jss.2024.111981},
	pages = {111981},
	journal = {Journal of Systems and Software},
	author = {Das, Souvick and Deb, Novarun and Cortesi, Agostino and Chaki, Nabendu},
}

@article{akbar2023ethical,
  author={Akbar, Muhammad Azeem and Khan, Arif Ali and Liang, Peng},
  journal={IEEE Transactions on Artificial Intelligence}, 
  title={Ethical Aspects of ChatGPT in Software Engineering Research}, 
  year={2025},
  volume={6},
  number={2},
  pages={254-267},
  doi={10.1109/TAI.2023.3318183}
}

@article{bozyigit2024generating,
  title={Generating domain models from natural language text using NLP: a benchmark dataset and experimental comparison of tools},
  author={Bozyigit, Fatma and Bardakci, Tolgahan and Khalilipour, Alireza and Challenger, Moharram and Ramackers, Guus and Babur, {\"O}nder and Chaudron, Michel RV},
  journal={Software and Systems Modeling},
  volume={23},
  number={6},
  pages={1493--1511},
  year={2024},
  publisher={Springer}
}

@misc{siddeshwar2024comparative,
    author = {Siddeshwar, Vaishali and Alwidian, Sanaa and Makrehchi, Masoud},
    year = {2024},
    month = {10},
    pages = {253-263},
    title = {A Comparative Study of Large Language Models for Goal Model Extraction},
    doi = {10.1145/3652620.3686246}
}

@misc{andrea_arcuri_webfuzzingemb_2025,
    title = {{WebFuzzing}/{EMB}: v3.4.0},
    copyright = {Apache License 2.0},
    shorttitle = {{WebFuzzing}/{EMB}},
    url = {https://zenodo.org/doi/10.5281/zenodo.14597431},
    urldate = {2026-01-02},
    publisher = {Zenodo},
    author = {Andrea Arcuri and Man Zhang and Amid Golmohammadi and Asma Belhadi and Onur Duman and Susruthan Seran and Juan Pablo Galeotti and Hernan Ghianni},
    month = jan,
    year = {2025},
    doi = {10.5281/ZENODO.14597431},
}

@misc{lin_how_2023,
    title = {How to write effective prompts for large language models},
    copyright = {https://creativecommons.org/licenses/by/4.0/legalcode},
    url = {https://osf.io/r78fc_v1},
    doi = {10.31234/osf.io/r78fc},
    urldate = {2026-01-04},
    publisher = {PsyArXiv},
    author = {Lin, Zhicheng},
    month = sep,
    year = {2023},
}

@inproceedings{errica_what_2025,
    address = {Albuquerque, New Mexico},
    title = {What {Did} {I} {Do} {Wrong}? {Quantifying} {LLMs}’ {Sensitivity} and {Consistency} to {Prompt} {Engineering}},
    shorttitle = {What {Did} {I} {Do} {Wrong}?},
    url = {https://aclanthology.org/2025.naacl-long.73},
    doi = {10.18653/v1/2025.naacl-long.73},
    language = {en},
    urldate = {2026-01-04},
    booktitle = {Proceedings of the 2025 {Conference} of the {Nations} of the {Americas} {Chapter} of the {Association} for {Computational} {Linguistics}: {Human} {Language} {Technologies} ({Volume} 1: {Long} {Papers})},
    publisher = {Association for Computational Linguistics},
    author = {Errica, Federico and Sanvito, Davide and Siracusano, Giuseppe and Bifulco, Roberto},
    year = {2025},
    pages = {1543--1558},
}

@misc{sami_ai_2024,
    title = {{AI} based {Multiagent} {Approach} for {Requirements} {Elicitation} and {Analysis}},
    url = {https://researchportal.tuni.fi/en/publications/ai-based-multiagent-approach-for-requirements-elicitation-and-ana/},
    doi = {10.48550/arXiv.2409.00038},
    language = {English},
    urldate = {2026-01-21},
    author = {Sami, Malik Abdul and Waseem, Muhammad and Zhang, Zheying and Rasheed, Zeeshan and Systä, Kari and Abrahamsson, Pekka},
    year = {2024},
}

@misc{alliance_what_2015,
    title = {What does {INVEST} {Stand} {For}? {\textbar} {Agile} {Alliance}},
    shorttitle = {What does {INVEST} {Stand} {For}?},
    url = {https://agilealliance.org/glossary/invest/},
    language = {en-US},
    urldate = {2026-01-21},
    author = {Alliance, Agile},
    month = dec,
    year = {2015},
}

@misc{zadenoori_large_2025,
    title = {Large {Language} {Models} ({LLMs}) for {Requirements} {Engineering} ({RE}): {A} {Systematic} {Literature} {Review}},
    shorttitle = {Large {Language} {Models} ({LLMs}) for {Requirements} {Engineering} ({RE})},
    url = {http://arxiv.org/abs/2509.11446},
    doi = {10.48550/arXiv.2509.11446},
    urldate = {2026-01-21},
    publisher = {arXiv},
    author = {Zadenoori, Mohammad Amin and Dąbrowski, Jacek and Alhoshan, Waad and Zhao, Liping and Ferrari, Alessio},
    month = sep,
    year = {2025},
    note = {arXiv:2509.11446 [cs]},
}

\end{document}